\documentclass[review]{elsarticle}
\usepackage{amssymb}

\newcommand{\mbfx}{\mathbf{x}}

\newcommand{\btPsi}{\bar{\tilde{\Psi}}}
\newcommand{\be}{\begin{equation}}
\newcommand{\ee}{\end{equation}}
\newcommand{\clF}{{\cal F}}
\newcommand{\clhH}{\hat{\cal H}} %\mathcal{}

\newcommand{\clL}{{\cal L}}
\newcommand{\hH}{\hat{H}}
\newcommand{\bfx}{{\bf x}}

\newcommand{\clW}{{\cal W}}

\newcommand{\bea}{\begin{eqnarray}}
\newcommand{\eea}{\end{eqnarray}}
\newcommand{\hh}{\tilde{h}}
\newcommand{\prt}{\partial}

\newcommand{\rgl}{\rangle}
\newcommand{\lgl}{\langle}

\begin{document}

\begin{frontmatter}

\title{Fractional-Time Schr\"odinger Equation: Fractional Dynamics on a
Comb}
\author{Alexander Iomin}
\ead{iomin@physics.technion.ac.il}
\address{Department of Physics, Technion, Haifa, 32000, Israel}

\begin{abstract}
The physical relevance of the fractional time derivative in quantum
mechanics is discussed. It is shown that the introduction of the
fractional time Scr\"odinger equation (FTSE) in quantum mechanics
by analogy with the fractional diffusion
$\frac{\prt}{\prt t}\rightarrow \frac{\prt^{\alpha}}{\prt t^{\alpha}}$
can lead to an essential
deficiency in the quantum mechanical description, and needs special
care. To shed light on this situation, a quantum comb model is
introduced. It is shown that for $\alpha=1/2$, the FTSE is a
particular case of the quantum comb model. This
\textit{exact} example shows that the FTSE is insufficient to
describe a quantum process, and the appearance of the fractional
time derivative by a simple change $\frac{\prt}{\prt t}\rightarrow
\frac{\prt^{\alpha}}{\prt t^{\alpha}}$ in the Schr\"odinger
equation leads to the loss of most of the information about quantum
dynamics.

\end{abstract}


\end{frontmatter}

\section{Introduction}

In quantum physics, the fractional concept can be introduced by
means of the Feynman propagator for non-relativistic quantum
mechanics as for Brownian path integrals \cite{feynman}.
Equivalence between the Wiener and the Feynman path integrals,
established by Kac \cite{kac}, indicates some relation between the
classical diffusion equation and the Schr\"odinger equation.
Therefore, the appearance of the space fractional derivatives in
the Schr\"odinger equation is natural, since both the standard
Schr\"odinger equation and the space fractional one obey the
Markov process. As shown in the seminal papers \cite{laskin1,
west}, it relates to the path integrals approach.  As a result of
this, the path integral approach for L\'evy stable processes,
leading to the fractional diffusion equation, can be extended to a
quantum Feynman-L\'evy measure which leads to the space fractional
Schr\"odinger equation \cite{laskin1,west}.

The fractional time Schr\"odinger equation (FTSE) was first considered in
\cite{naber}, where a fractional time derivative was introduced in
the quantum mechanics by analogy with the fractional Fokker-Planck
equation (FFPE),  by means of the Wick rotation of time
$t\rightarrow -it/\hbar$ . The dynamics does not correspond to
the unitary transformation: the
Green function  is found in the form of the Mittag-Leffler
function and does not satisfy Stone's theorem on one-parameter
unitary groups \cite{stone}.
A generalization for the space-time
fractional quantum dynamics \cite{DongXu,WangXu} was performed and
a relation to the fractional uncertainty \cite{bhatti} was studied
as well. It was also shown that the FTSE introduces new
nonlinear phenomena in the semiclassical limit, and this
semiclassical approach differs from those described in the
framework of the standard Schr\"odinger equation \cite{iom2009}.

The fractional time quantum dynamics with the Hamiltonian
$\hat{H}(x)$ is described by the FTSE
\begin{equation}\label{fse1}
(i\hh)^{\alpha}\frac{\prt^{\alpha}\psi(\bfx,t)}{\prt
t^{\alpha}}= \clhH\psi(\bfx,t)\, , %
\end{equation}
where $\alpha\leq 1$. For agreement of the dimension in Eq.
(\ref{fse1}), all variables and parameters are considered
dimensionless, and $\hh$ is the dimensionless Planck constant, see
also \cite{naber,DongXu}. For $\alpha=1$, Eq. (\ref{fse1}) is the
"conventional" (standard) Schr\"odinger equation. For $\alpha<1$
the fractional derivative is a formal notation of an integral with
a power law memory kernel of the form
\begin{equation}\label{fse2}
\frac{\prt^{\alpha}\psi(t)}{\prt t^{\alpha}}\equiv
I_t^{1-\alpha}\frac{\prt\psi(t)}{\prt t}
=\int_0^t\frac{(t-\tau)^{-\alpha}}{\Gamma(1-\alpha)}
\frac{\prt\psi(\tau)}{\prt\tau}d\tau\, , %
\end{equation} %
which is the Caputo fractional derivative:
$\frac{\prt^{\alpha}\psi(t)}{\prt t^{\alpha}}\equiv
{}_0D_{C}^{\alpha}\psi(t)$ (see Appendix).

In this paper we present an \textit{exact} example where the
fractional time derivative is naturally introduced and has a well
defined physical meaning. We consider a case with $\alpha=1/2$,
when fractional quantum dynamics can be modelled by means of the
conventional quantum mechanics in the framework of a comb model,
and the FFPE for $\alpha=1/2$ is a particular case of the quantum
comb model which is a quantum counterpart of a diffusive comb
model \cite{em1,bi2004}. We also show that the FTSE (\ref{fse2})
is insufficient to describe a quantum process, and the appearance
of the fractional time derivative leads to the loss of most of the
information about quantum dynamics. The main idea is to show that
Eq. (\ref{fse1}) is a result of a projection of the
two-dimensional $(x,y)$ comb dynamics in the one-dimensional
configuration space. For diffusion, which is described by the
Fokker-Planck equation, this projection is a simple integration
over the $y$ space. In quantum mechanics this projection is
performed by means of the Fourier transform
$\Psi(x,y)\rightarrow\bar{\Psi}_l(x)={\cal F}_y\Psi(x,y)$, where
$l$ is the Fourier index. For the comb model this procedure can be
treated exactly, and we shall show that Eq. (\ref{fse1}) is valid
only for the zero Fourier component
$\psi(x)\equiv\bar{\Psi}_0(x)$. All other components are not
described by the FTSE (\ref{fse1}) and lost in the framework of
this equation. The diffusive comb model is an analogue of a 1d
medium where fractional diffusion has been observed
\cite{em1,bi2004}. It is a particular example of a non-Markovian
phenomenon, explained in the framework of the so-called continuous
time random walks (CTRW) \cite{em1,shlesinger,klafter}. This model
is also known as a toy model for a porous medium used for
exploring of low dimensional percolation clusters \cite{baskin1}.

\section{Quantum Comb Model and FTSE}

A special quantum behavior of a particle on the comb can be
defined as the quantum motion in the $d+1$ configuration space
$(\mathbf{x},y)$, such that the dynamics in the $d$ dimensional
configuration space $\mathbf{x}$ is possible only at $y=0$, and
motions in the $\mathbf{x}$ and $y$ directions commute. Therefore
the quantum dynamics is described by the following Schr\"odinger
equation
\begin{equation}\label{qcomb1}%
i\hh\frac{\prt\Psi}{\prt t}=\delta(y)\hat{H}(\mathbf{x})\Psi-
\frac{\hh^2}{2}\frac{\prt^2\Psi}{\prt y^2}\, , %
 \end{equation} %
where the Hamiltonian
$\hH\equiv\hat{H}(\mathbf{x})=-\frac{\hh^2}{2}\nabla+V(\mathbf{x})$
can be different from $clhH$ in Eq. (\ref{fse1}). It governs the dynamics
with a potential $V(\mathbf{x})$ in the $\mathbf{x}$ space, while
the $y$ coordinate corresponds to the 1d free motion. All the
parameters and variables are dimensionless \footnote{Analogously
to the FTSE (\ref{fse1}), following Ref. \cite{naber}, one
introduces the Planck length $L_P=\sqrt{\hbar G/c^3}$, time
$T_P=\sqrt{\hbar G/c^5}$, mass $M_P=\sqrt{\hbar c/G}$, and energy
$E_P=M_Pc^2$, where $\hbar,~G,$ and $c$ are the Planck constant,
the gravitational constant and the speed of light, respectively.
Therefore, quantum mechanics of a particle with mass $m$ is
described by the dimensionless units $x/L_P\rightarrow
x,~y/L_P\rightarrow y,~ t/T_P\rightarrow t$, while the
dimensionless Planck constant is defined as the inverse
dimensionless mass $\hh=M_P/m$. Note, that the dimensionless
potential is now $V(\bfx)\rightarrow V(\bfx)/M_Pc^2$.}.

By analogy with the diffusion (classical) comb model, we are concerned
with the dynamics in the $\mathbf{x}$ space. But simple
integration of the wave function over the $y$ coordinate is not
valid. Therefore, one carries out the Fourier transform in
the $y$ space
$\clF_y\Psi(\mathbf{x},y,t)=
\bar{\Psi}(\mathbf{x},l,t)\equiv\bar{\Psi}_l$,
and as a result of this, Eq. (\ref{qcomb1}) reads
\begin{equation}\label{qcomb2}   %
i\hh\frac{\prt\bar{\Psi}_l}{\prt t}=
\hat{H}(\mathbf{x})\Psi(\mbfx,0,t)+\frac{\hh l^2}{2}\bar{\Psi}_l\, .
\end{equation}   %
To obtain this equation in a closed form, one needs to express the
wave function at $y=0$ $\Psi(\mbfx,0,t)$ by the Fourier image
$\bar{\Psi}_l$. To this end the Laplace transform of Eq.
(\ref{qcomb1}) is performed
$\clL[\Psi(\mbfx,y,t)]=\tilde{\Psi}_s(\mbfx,y)$. The
solution in the Laplace domain reads
\begin{equation}\label{qcomb3}   %
\tilde{\Psi}_s(\mbfx,y)=
\bar{\Psi}_s(\mbfx,0)\exp\Big[i(1+i)\sqrt{s/\hh}|y|\Big]\,
, \end{equation}  %
where we used $\sqrt{2i}=(1+i)$. Performing the Fourier transform
 $\btPsi_{s,l}(\mbfx)=\clF[\tilde{\Psi}_s(\mbfx,y)]$ one obtains
from Eq. (\ref{qcomb3})
\begin{equation}\label{qcomb4}    %
\btPsi_{s,l}(\mbfx)=\bar{\Psi}_s(\mbfx,0)\clF_ye^{i(1+i)\sqrt{s/\hh}|y|}
= \frac{2i(1+i)\sqrt{s/\hh}}{l^2-2is/\hh}\bar{\Psi}_s(\mbfx,0)\,
.\end{equation} %
Finally, the Laplace inversion for $\bar{\Psi}_s(\mbfx,0)$
determines the wave function at $y=0$
\begin{equation}\label{qcomb5}   %
\Psi(\mbfx,0,t)=\clL^{-1}
\Big[\btPsi_{s,l}(\mbfx)\frac{l^2-2is/h}{2i(1+i)\sqrt{s/\hh}}\Big]\,
.\end{equation}     %
Let us, first, consider a simple case with $l=0$. We have from Eq.
(\ref{qcomb5}) $\tilde{\Psi}_s(\mbfx,0)=-\sqrt{\frac{s}{2\hh
i}}\btPsi_{s,0}$. Then we define $\bar{\Psi}_0(\mbfx,t)=\psi(x)$,
and, carrying out the Laplace transform in Eq. (\ref{qcomb2}), we
arrive at the definition of the Caputo fractional derivative
(\ref{fse2}) in the Laplace domain (see Appendix)
$\clL[{}_0D_t^{1/2}\psi(t)]=s^{1/2}\tilde{\psi}(s)-s^{-1/2}\psi(0)$.
Finally, carrying out the inverse Laplace transform and redefining
$\frac{\hat{iH}}{\sqrt{2}\hh}\rightarrow\clhH$ , one obtains
the FTSE which coincides exactly with Eq. (\ref{fse1}) for
$\alpha=1/2$.

Repeating  this procedure for an arbitrary $l$, one
performs the Laplace transform of the term proportional to
$l^2/\sqrt{s}$ in Eq. (\ref{qcomb5}). Performing simple operations
of fractional calculus and taking into account Eqs. (\ref{mt1a})
and (\ref{mt3}), one obtains

\begin{equation}\label{qcomb7}    %
(i\hh)^{\frac{1}{2}}
\frac{\prt^{\frac{1}{2}}\bar{\Psi}_l}{\prt t^{\frac{1}{2}}}=
-\frac{l^2}{2\sqrt{2}}\,{}_0I_t^{1}\hat{H}(\mathbf{x})\bar{\Psi}_l
+\frac{i}{\sqrt{2}\hh}
\hat{H}\bar{\Psi}_l+\frac{\hh^2l^2}{2}\bar{\Psi}_l\, .
\end{equation}    %
This comb FTSE describes the quantum dynamics in the $\mbfx$
configuration space. The index $l$ corresponds to an effective
interaction of a quantum system with an additional degree of
freedom, while the fractional time derivatives, with $\alpha=1/2$,
reflect this interaction in the form of non-Markov memory effects.

\section{Green's Function}

The initial value problem with the initial condition
$\Psi(t=0)=\Psi_0(\mathbf{x},y)$ is described by Green's
function. For the complete analogy with the classical comb model
and fractional diffusion \cite{bi2004,ib2005}, the boundary
conditions for the $y$ direction defined at infinities
$y=\pm\infty$ are $\Psi(\bfx,t)=\prt\Psi(\bfx,t)/\prt y=0$. Using
the eigenvalue problem
 \begin{equation}\label{qcom1_a}%
\hat{H}(\mathbf{x})\psi_{\lambda}(\bfx)=\lambda\psi_{\lambda}(\bfx)\,
,\end{equation} %
we present the wave function in Eq. (\ref{qcomb1}) as the
expansion
$\Psi(\bfx,y,t)=\sum_{\lambda}\phi_{\lambda}(y,t)\psi_{\lambda}(\bfx)$,
where $\sum_{\lambda}$ also supposes integration on $\lambda$ for
the continuous spectrum. For the fixed $\lambda$ we arrive at the
dynamics of a particle in the $\delta$ potential
\begin{equation}\label{qcomb8} %
 i\hh\frac{\prt\phi_{\lambda}}{\prt
t}=-\frac{\hh^2}{2}\frac{\prt^2\phi_{\lambda}}{\prt
y^2}+\lambda\delta(y)\phi_{\lambda} \, . %
\end{equation} %
Taking into account that the Green function of Eq. (\ref{qcomb1})
has the spectral decomposition,
\[G(\bfx,y,t;\bfx'y')=\sum_{\lambda}G_{\lambda}(y,t;y')
\psi_{\lambda}^*(\bfx)\psi_{\lambda}(\bfx')\, ,\] we obtain that
the Schr\"odinger equation for the Green function
$G_{\lambda}(y,t;y')$ reads
\begin{equation}\label{qcomb9} %
 i\hh\frac{\prt G_{\lambda}}{\prt
t}=-\frac{\hh^2}{2}\frac{\prt^2G_{\lambda}}{\prt
y^2}+\lambda\delta(y)G_{\lambda}+i\hh\delta(y)\delta(t) \, . %
\end{equation} %
Here the initial condition is already taken into account.
The Green function for this Schr\"odinger equation has been
obtained in \cite{Lary2,Lary1}, for free boundary conditions at
infinities. For the chosen boundary conditions it is instructive
to employ Eq. (\ref{qcomb3}) in the Laplace domain. Then replacing
the eigenvalues $\lambda$ by the Hamiltonian
$\hat{H}(\mathbf{x})$, one obtains the Green's function in the
form of the inverse Laplace transform
\begin{equation}\label{qcomb10}   %
G\Big[\hat{H}(\bfx),y,t)\Big]=\clL^{-1}\Big[\frac{i\hh
e^{i(1+i)\sqrt{s/\hh}|y|}}{\hat{H}(\bfx)-i(1+i)\sqrt{\hh^3s}}\Big]\,
. \end{equation} %
One performs the inverse Laplace transform, using the following
presentation for the denominator
\[\int_0^{\infty}\exp\Big\{-u[\hat{H}(\bfx)-i(1+i)\sqrt{\hh^3s}]\Big\}
du\, .\]
This presentation is valid for any spectrum $\lambda$ due to the
second term in the exponential.
Therefore, the Green function reads
\begin{equation}\label{qcomb11}   %
G\Big[\hat{H}(\bfx),y,t)\Big]= \frac{\sqrt{i\hh}}{\sqrt{2\pi t^3}}
\int_0^{\infty}(|y|+\hh^2u)
\exp\Big[-u\hat{H}(\bfx)-\frac{i(|y|+\hh^2u)^2}{2\hh t}\Big]du\, .
\end{equation}
Using the Fourier transform for the exponential
\[\exp[\frac{i(|y|+\hh^2u)^2}{2\hh t}]=\sqrt{\frac{\hh t}{2\pi i}}
\int_{-\infty}^{\infty}e^{i\hh t\xi^2/2}e^{-i\xi(|y|+\hh^2
u)}d\xi\, ,\] one presents the Green function in the following
convenient form
\begin{eqnarray}\label{PI1}
G(x,y,t;x')&=&\frac{h}{2\pi t} \int_{-\infty}^{\infty} \left\{
e^{i\hh t\xi^2/2}\Big(i\frac{\prt}{\prt\xi}\Big)\right.
\nonumber \\
&\times& \left.\int_0^{\infty}
\exp\Big[-u\hat{H}(\bfx)-i\xi(|y|+\hh^2 u)\Big] du\right\}d\xi\, .
\end{eqnarray}
This expression is convenient for further analysis in the
framework of the path integral.

\section{Path Integral Presentation}

As an example, it is instructive to consider a free particle,
because it has a straightforward relation to the original
diffusive comb model \cite{em1,bi2004}. Therefore, we find the
Green function along the structure $x$ axis in the coordinate
space $G(x,y,t;x')=\lgl x'|G\Big[\hat{H}(\bfx),y,t\Big]|x\rgl$ for
a free particle of a unit mass with the Hamiltonian
$\hat{H}={p}^2/2$.  Expressing the exponential $\lgl
x'|e^{-u\hat{H}}|x\rgl$ in the path integral form,
$\frac{1}{\sqrt{2\pi\hh^2u}}e^{-\frac{(x-x')^2}{2u\hh^2}}$, one
obtains from Eq. (\ref{PI1}):
\begin{equation}\label{PI1a}
G(x,y,t;x')=\frac{1}{t\sqrt{(2\pi)^3}}
\int_{-\infty}^{\infty}\left\{
e^{i\hh t\xi^2/2}\Big(i\frac{\prt}{\prt\xi}\Big)
\int_0^{\infty}
\exp\Big[-\frac{(x-x')^2}{2\hh^2u}-i\xi(|y|+\hh^2 u)\Big]
\frac{du}{\sqrt{u}}\right\}d\xi\, .
\end{equation}
Integration over complex ``time'' $u$ yields the following
expression
\begin{equation}\label{PI2}
I(\mathcal{A},\mathcal{B})=\int_0^{\infty}
\exp[-\frac{\mathcal{A}}{u}-\mathcal{B}u]
\frac{du}{\sqrt{u}}\, ,
\end{equation}
where $\mathcal{A}=\frac{(x-x')^2}{2\hh^2}$ and
$\mathcal{B}=-i\xi\hh^2$. Differentiation of Eq. (\ref{PI2}) with
respect to $\mathcal{A}$ and $\mathcal{B}$ yields the equation
$\frac{\partial^2I(\mathcal{A},\mathcal{B})}
{\partial\mathcal{B}\partial\mathcal{A}}
=I(\mathcal{A},\mathcal{B})$. A solution of this equation is
\[I(\mathcal{A},\mathcal{B})=
\sqrt{\frac{\pi}{4\sqrt{\mathcal{A}\mathcal{B}}}}
\exp[2\sqrt{\mathcal{A}\mathcal{B}}]\, .\] %
Performing this integration, we arrive at integration over $\xi$
that is carried out in the stationary phase approximation for the
long time asymptotics $\hh t\gg 1$. This yields
\begin{equation}\label{PI3}
\int_{-\infty}^{\infty}F(\xi)\exp[i\hh t\xi^2-i\xi|y|]\approx
\sqrt{\frac{\pi }{i\hh t}}F(\xi_0)\exp\Big[-i\frac{y^2}{4\hh
t}\Big]\, ,
\end{equation}
where the stationary point is $\xi_0=\frac{|y|}{2\hh t}$. Taking
integration (\ref{PI3}) into account, we finally, obtain the Green
function in the form
\begin{eqnarray}\label{PI4}
G(x,y,t;x')&\approx&\frac{i^{1/4}}{4\pi t\sqrt{2|x-x'|}}
\left[-\frac{i}{4}\left(\frac{|u|}{\hh
t}\right)^{-\frac{5}{4}}+|y|-|x-x'|\sqrt{\frac{i\hh
t}{2|y|}}\right] \nonumber  \\
 &\times&\exp \Big[ -i\frac{y^2}{2\hh
t}+i\sqrt{i|y|/\hh t}|x-x'|\Big]\, .
\end{eqnarray}
This solution also satisfies the boundary conditions at
$x,y=\pm\infty$, where the Green function vanishes.

\section{Conclusion}

Physical relevance of the fractional time derivative in quantum
mechanics  is discussed. It is shown that the introduction of the
fractional time Scr\"odinger equation in quantum mechanics by the
change  $\frac{\prt}{\prt t}\rightarrow \frac{\prt^{\alpha}}{\prt
t^{\alpha}}$ by an analogy with the fractional diffusion can lead
to essential deficiency of the quantum mechanical description, and
needs special care. To shed light on this situation, a quantum
comb model is introduced. We observed that the fractional time
derivative, at least for $\alpha=1/2$, reflects an effective
interaction of a quantum system with an additional degree of
freedom. In the classical diffusion comb model, diffusion in the
$y$ direction is responsible for traps that lead to subdiffusion
along the $x$ structure axis, and this phenomenon is described by
the time fractional derivative $\frac{\prt^{\frac{1}{2}}}{\prt
t^{\frac{1}{2}}}$. This description in the framework of the
fractional Fokker-Planck equation is identical to the diffusion
comb model \cite{bi2004,ib2005}. In the quantum case the situation
differs essentially from fractional diffusion. First of all, the
quantum comb model (\ref{qcomb1}) and the FTSE (\ref{fse1}) are
not identical. As shown here, the FTSE is an equation only for the
zero Fourier component of the wave function, and this equation is
insufficient to describe the complete dynamics in the $x$ space of
the system. The FTSE is a particular and limiting case of the comb
fractional time Schr\"odinger equation (\ref{qcomb7}), obtained
here, which describes the quantum dynamics in the $\mbfx$
configuration space, and the Fourier index $l$ corresponds to an
effective interaction of the quantum Hamiltonian $\hH(\bfx)$ with
an additional degree of freedom, while the fractional time
derivatives, with $\alpha=1/2$, reflect this interaction in the form
of non-Markov memory effects. This equation can be readily solved
by the Laplace transform and using the eigenvalue problem
$\hat{H}(\mathbf{x})\psi_{\lambda}(\bfx)=\lambda\psi_{\lambda}(\bfx)$.
Nevertheless, we obtained the Green function directly from the
quantum comb model of Eq. (\ref{qcomb1}) in a form suitable for
the paths integral presentation. An example of $\hH(\bfx)$ that
corresponds to a free particle is considered. Note also that this
expression for the Green function is also suitable for the
semiclassical treatment of more complicated Hamiltonian systems
$\hH(\bfx)$.

In conclusion, we admit that this \textit{exact} example shows
that the FTSE (\ref{fse1}) is insufficient to describe a quantum
process, and the appearance of the fractional time derivative by a
simple change $\frac{\prt}{\prt t}\rightarrow
\frac{\prt^{\alpha}}{\prt t^{\alpha}}$ in the Schr\"odinger
equation leads to loss of most of the information about quantum
dynamics.

This work was
supported in part by the Israel Science Foundation (ISF) and by
the US-Israel Binational Science Foundation (BSF).

\section*{Appendix: Fractional Calculus Tools}
\def\theequation{A. \arabic{equation}}
\setcounter{equation}{0}

Fractional derivation was developed as a generalization of integer
order derivatives and is defined as the inverse operation to the
fractional integral. Fractional integration of the order of
$\alpha$ is defined by the operator (see {\em e.g.},
\cite{klafter,podlubny,oldham} )
\begin{equation}\label{A1}
{}_aI_x^{\alpha}f(x)=
\frac{1}{\Gamma(\alpha)}\int_a^xf(y)(x-y)^{\alpha-1}dy\, ,
\end{equation}
where $\alpha>0,~x>a$ and  $\Gamma(z)$ is the Gamma
function. Therefore, the fractional derivative is defined as the
inverse operator to ${}_aI_x^{\alpha}$, namely $
{}_aD_x^{\alpha}f(x)={}_aI_x^{-\alpha}f(x)$ and
${}_aI_x^{\alpha}={}_aD_x^{-\alpha}$. Its explicit form is
\begin{equation}\label{A2}
{}_aD_x^{\alpha}f(x)=
\frac{1}{\Gamma(-\alpha)}\int_a^xf(y)(x-y)^{-1-\alpha}dy\, .
\end{equation}
For arbitrary $\alpha>0$ this integral diverges, and as a result
of this a regularization procedure is introduced with two
alternative definitions of ${}_aD_x^{\alpha}$. For an integer $n$
defined as $n-1<\alpha<n$, one obtains the Riemann-Liouville
fractional derivative of the form
\begin{equation}\label{mt1a}   %
{}_aD_{RL}^{\alpha}f(x)=\frac{d^n}{dx^n}{}_aI_x^{n-\alpha}f(x)\, ,
\end{equation}
and fractional derivative in the Caputo form (see also
\cite{mainardi})
\begin{equation}\label{mt1b}  %
{}_aD_{C}^{\alpha}f(x)= {}_aI_x^{n-\alpha}f^{(n)}(x)\, , ~~~
f^{(n)}(x)\equiv\frac{d^n}{dx^n}f(x)\, .
\end{equation}  %
There is no constraint on the lower limit $a$. For example, when
$a=0$, one has
\[{}_0D_{RL}^{\alpha}x^{\beta}=\frac{x^{\beta-\alpha}
\Gamma(\beta+1)}{\Gamma(\beta+1-\alpha)} \, .\] We also have
\[{}_0D_{C}^{\alpha}f(x)=
{}_0D_{RL}^{\alpha}f(x)-\sum_{k=0}^{n-1}f^{(k)}(0^+)
\frac{x^{k-\alpha}}{\Gamma(k-\alpha+1)}\, ,\] and
${}_aD_{C}^{\alpha}[1]=0$, while
${}_0D_{RL}^{\alpha}[1]=x^{-\alpha}/\Gamma(1-\alpha)$. When
$a=-\infty$, the resulting Weyl derivative is
\begin{equation}\label{weyldr1}
{}_{-\infty}\clW^{\alpha}\equiv{}_{-\infty}D_{W}^{\alpha}=
{}_{-\infty}D_{RL}^{\alpha}= {}_{-\infty}D_{C}^{\alpha}\, .
\end{equation}
One also has ${}_{-\infty}D_{W}^{\alpha}e^x=e^x$. This property is
convenient for the Fourier transform
\begin{equation}\label{top_0a}
\clF\left[{}_{-\infty}\clW^{\alpha}f(x)\right]=(ik)^{\alpha}\bar{f}(k)\, ,  %
\end{equation}  %
where $\clF[f(x)]=\bar{f}(k)$. This fractional derivation with the
fixed low limit is also called the left fractional derivative.
However, one can introduce the right fractional derivative, where
the upper limit $a$ is fixed and $a>x$. For example, the right
Weyl derivative is
\begin{equation}\label{top_0b}
\clW_{\infty}^{\alpha}f(z)=\frac{1}{\Gamma(-\alpha)}
\int_x^{\infty}\frac{f(y)dy}{(y-x)^{1+\alpha}}\, .
\end{equation}  %
The Laplace transform of the Caputo fractional derivative yields
\begin{equation}\label{mt2}
 \clL[{}_0D_{C}^{\alpha}f(x)]=
s^{\alpha}\tilde{f}(s)-\sum_{k=0}^{n-1}f^{(k)}(0^+)s^{\alpha-1-k}
\, ,
\end{equation}
where $ \clL[f(x)]=\tilde{f}(s)$, which is convenient for the
present analysis, where the initial conditions are imposed in
terms of integer derivatives. We also use here a convolution rule
for $0<\alpha<1$
\begin{equation}\label{mt3}
\clL[{}I_x^{\alpha}f(x)]=s^{-\alpha}\tilde{f}(s)\, .
\end{equation}

\end{document}